\documentclass[a4paper, 11pt]{article}

\usepackage[affil-it]{authblk}

\makeatletter
\def\@maketitle{%
  \newpage
  \null
  \vskip 2em
  \begin{center}%
  \let \footnote \thanks
    {\Large\bfseries \@title \par}%
    \vskip 1.5em
    {\normalsize
      \lineskip .5em %
      \begin{tabular}[t]{c}%
         \@author
      \end{tabular} \par}%
    \vskip 1em 
    {\normalsize \@date}%
  \end{center}%
  \par
  \vskip 1.5em} 
\makeatother

\usepackage{AdamStyle}
\usepackage{amsmath}

\providecommand{\keywords}[1]{\textbf{\textit{\upshape Keywords:}} #1}

\newcommand{\nP}{\mathcal{P}}
\newcommand{\nV}{\mathcal{V}}
\newcommand{\MS}{{\rm MS}}
\newcommand{\dP}{{d\!P}}

\title{Increased blood pressure variability upon standing up improves reproducibility of cerebral autoregulation indices}

\author{A \!Mahdi$^{\,a}$, D \!Nikolic$^{\,b}$, AA \!Birch$^{\,c}$, MS Olufsen$^{\,d}$, RB \!Panerai$^{\,e}$, DM \!Simpson$^{\,b}$, SJ \!Payne$^{\,a}$}

{
\affil{$^{a\,}$\small Institute of Biomedical Engineering, Department of Engineering Science, \\
University of Oxford, Oxford, UK}

\affil{$^{b\,}$Institute of Sound and Vibration Research, \\University of Southampton, Southampton, UK}

\affil{$^{c\,}$ Department of Medical Physics and Bioengineering, \\Southampton General Hospital, Southampton SO16 6YD, UK} 

\affil{$^{d\,}$ Department of Mathematics, \\North Carolina State University, Raleigh, USA} 

\affil{$^{e\,}$Department of Cardiovascular Sciences \\and NIHR Cardiovascular Biomedical Research Unit, \\University of Leicester, Leicester, UK} 
}
\date{}

\newcommand{\argmin}{\operatornamewithlimits{arg\,\,min}}
\newcommand{\ari}{{\rm ARI}}

\begin{document}

\maketitle

\bigskip

\vspace{-1cm}
\begin{abstract}
{\it Background:} Dynamic cerebral autoregulation, that is the transient response of cerebral blood flow to changes in arterial blood pressure, is currently assessed using a variety of different time series methods and data collection protocols. In the continuing absence of a gold standard for the study of cerebral autoregulation it is unclear to what extent does the  assessment depend on the choice of a computational method and protocol.
{\it Methods:} We use continuous measurements of blood pressure and cerebral blood flow velocity in the middle cerebral artery from the cohorts of 18 normotensive subjects performing sit-to-stand manoeuvre. We estimate cerebral autoregulation using a wide variety of black-box approaches (ARI, Mx, Sx, Dx, FIR and ARX)  and compare them in the context of reproducibility and variability.
{\it Results:} For all autoregulation indices, considered here, the ICC was greater during the standing protocol, however, it was significantly greater (Fisher's Z-test) for  Mx ($p < 0.03$), Sx $(p<0.003)$ and Dx ($p<0.03$). 
 {\it Conclusions:} In the specific case of the sit-to-stand manoeuvre,  measurements taken immediately after standing up greatly improve the reproducibility of the autoregulation coefficients. This is generally coupled with an increase of the within-group spread of the estimates. 
\end{abstract}


\onehalfspacing

\section{Introduction}

Cerebral autoregulation (CA) refers to the brain's control mechanisms responsible for maintaining cerebral blood flow at an appropriate, approximately constant, level despite changes in arterial blood pressure (ABP) \cite{Aaslid1989}. Lassen \cite{Lassen59} was the first to show this phenomenon, by plotting the so-called {autoregulation curve} combining the measurements from different human studies \cite{Dagal2009}. Other authors have obtained similar results both in animals \cite{MacKenzie76, MacKenzie79, Harper84, Dirnagl90} and more recently in humans \cite{Strandgaard1976, Czosnyka2001}.

The application of Doppler ultrasound in obtaining continuous measurements of cerebral blood flow velocity (CBFV) in the middle cerebral artery (MCA) has allowed the study of cerebral blood flow noninvasively (under the assumption that the vessel diameter remains constant). This has stimulated the development of quantitative methods for CA assessment. In particular, it has allowed the study of dynamic aspects of CA by considering the adaptation of CBFV in response to ABP change.

Having fast, reliable and noninvasive autoregulation assessment techniques is of great importance because of the link between CA impairment and many clinical disorders. For example, poor CA assessment has been demonstrated in  stroke \cite{Dawson2000}, subarachnoid hemorrhage \cite{Giller1990} and head injury \cite{Czosnyka1996}. Other studies have also pointed to a potential link between impaired autoregulation and syncope or cerebral microvascular disease \cite{Novak2006}. 

Over the years many different mathematical methods have been developed in order to quantify autoregulation. Common approaches include transfer function analysis (TFA) \cite{Claassen2016}, autoregulation index (ARI) \cite{Tiecks95}, parametric time series models such as FIR \cite{Panerai2000} and ARX \cite{Liu02, Liu03}, the correlation indices Mx, Sx and Dx \cite{Piechnik1999, Reinhard2003}, neural networks \cite{Panerai2004}; wavelet synchronization analysis \cite{Latka2005, Peng2010} and other modelling techniques \cite{UL97, UL00, Payne06, Mader2015}. Experimental protocols have been designed that focus on spontaneously occurring variability in ABP-CBFV signals and those that induce changes in ABP using different manoeuvres. The latter include lower body negative pressure \cite{Birch2001}, thigh-cuffs \cite{Mahony2000}, controlled breathing  \cite{Panerai03}, cyclic leg raising  \cite{Elting2014}, and sit-to-stand \cite{Lip2000} maneuvers. One aspect that has been noted is that increased variability of blood pressure leads  to more robust estimates of autoregulation (see \cite{Simpson2004,Liu2005}).

The vast majority of studies use only one or a few measures of autoregulation for a given data collection protocol. In the continuing absence of a gold standard the choice of a computational method to assess dynamic CA  is not obvious. In practice, the choice is often ad-hoc or based on the authors' preference in using a specific  technique. Despite the fast growing body of literature on CA it remains unclear what the difference is between various quantitative methods and their dependence on the data collection protocols.

In Angarita-Jaimes et al. \cite{Angarita-Jaimes2014} the authors explored a number of different autoregulation parameters and compared them in normocapnia and hypercapnia (which is known to impair autoregulation)  using Monte-Carlo simulations to assess within-subject measurement errors. Lower between- and within-subject variability of the parameters were considered as criteria for identifying an improved metric of CA.  In another recent study  Nikolic et al. \cite{Nikolic2016} compared the reproducibility  of CA measures based on the phase and gain of FIR and IIR filters of many different orders (0-20) for the baseline and thigh-cuff manoeuvres. 

The current paper extends the previous works by considering the short-term repeatability of autoregulation measures considering repeatability within the same recording session, and comparing rest with standing-up, as a maneuver that increases blood pressure variations. We use six types of black-box approaches (ARI, Mx, Sx, Dx, FIR and ARX) for estimating autoregulaton in the context of reproducibility and variability for two different protocols: baseline (sitting) and orthostatic stress (sit-to-stand manoeuvre).  Although some authors have studied various aspects of autoregulation during sit-to-stand, the assessment was based primarily on a few specific CA coefficients such as ARI or TFA \cite{Beek2010, Lip2000}. As far as we know, this is one of the first studies that compares such a  wide range of different autoregulation coefficients on the same dataset with repeated measurements.  

\section{Methods}\label{sec:Methods}
In this section we describe the data collection protocol, data preprocessing and computational methods for CA assessment.

\subsection{Data collection protocol}\label{data}
\noindent {\it Data collection.}
The ABP and CBFV data, used in this study, have been previously discussed in Lipsitz et al.~\cite{Lip2000}. ABP was measured noninvasively using a photoplethysmographic Finapres monitor (Ohmeda Monitoring Systems, Englewood, CO). In order to eliminate hydrostatic pressure effects, the subject's nondominant hand was supported by a sling at the level of the right atrium. The individuals were asked to breathe at the rate of 15 breaths per minute to standarise the effects of respiration. Doppler ultrasonography was used to measure the changes in CBFV within the MCA. The 2 MHz probe of a portable Doppler system (MultiDop X4, DWL-Transcranial Doppler Systems Inc., Sterling, VA) was strapped over the temporal bone and locked in position with a Mueller-Moll probe fixation device. Flow velocity was recorded at a depth of approximately 50-65 mm, digitized and stored for analysis. After instrumentation, subjects sat in a straight-backed chair with their legs elevated at 90 degrees in front of them. First subjects rested in the sitting position for 5 minutes, then stood upright for one minute. The data were recorded during the final one minute of sitting and first minute of standing and the initiation of standing was timed from the moment both feet touched the floor. The active stand protocol was repeated in each subject. For a more detailed description of the data collection procedure see ~\cite{Lip2000}.

\smallskip

\noindent {\it Data preprocessing.}
Artifacts including spikes that commonly occur in CBFV signals were removed as the first step of data preprocessing using a median filter. The pulsatile ABP and CBFV were low-pass filtered using a 4th-order Butterworth filter, in both the forward and reverse directions, with a cutoff frequency of $20\,\mathrm{Hz}$ (see \cite{Panerai2000}). Subsequently, the beginning and end of each cardiac cycle were marked by the onset of the systole using the ABP signal. The onsets were detected using a windowed and weighted slope sum function and adaptive thresholding \cite{Zong2003}. The beat-to-beat average of ABP and CBFV were calculated for each detected cardiac cycle. A first-order polynomial was used to interpolate the resulting time series, which was followed by downsampling at $10\,\mathrm{Hz}$ to produce signals with a uniform time base. Preprocessed ABP and CBFV time series are denoted by $P[k]$ and $V[k]$, respectively, as used in each of the modelling approaches described below.

\smallskip

\begin{figure}[h!]
\begin{center}
\includegraphics[width=0.9\textwidth]{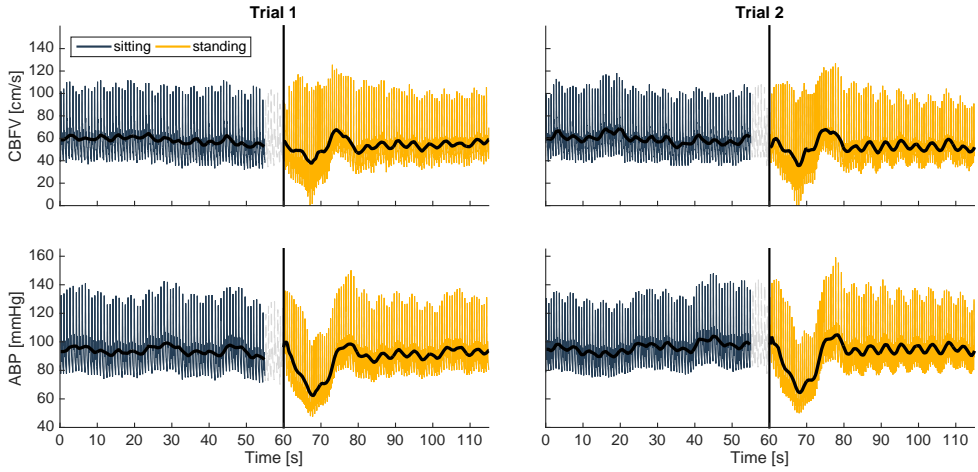}
\end{center}
\caption{An example of repeated measurement of ABP and CBFV from one normotensive individual during sit-to-stand exercise. The vertical line indicates the initiation of standing when both feet touched the floor. The beat-to-beat average of ABP and CBFV are shown with thick black lines. The gray dotted line shows 5 seconds of time series excluded from the analysis.}\label{fig:data}
\end{figure}

\noindent {\it Data segments for sitting and standing.}
To study autoregulation during the standing protocol approximately 55\,s data segments have been selected starting from the moment subjects stood up (see Fig.~\ref{fig:data}). The characteristic `dip', the overshoot and adaptation of the ABP-CBFV signals lasted approximately 20-30 seconds, depending on the individual's cardiovascular properties.  From the available 1 minute recording of the signals during the sitting protocol we removed the last 5 seconds of data in order to separate the baseline from the transient part of ABP-CBFV.

\subsection{Cerebral autoregulation estimates}

\noindent {\it ARI.}
Tiecks et al. \cite{Tiecks95} proposed  by of difference equations for CBFV response to a change in ABP, from which an autoregulation index (ARI) can be calculated.   The method itself and its variations have been used extensively to provide a quantitative assessment of CA  \cite{Panerai08, Panerai08b, Angarita-Jaimes2014}. The index  ranges from 0, representing the absence of autoregulation, to 9, indicating the best autoregulation. For more details on the computational aspects of ARI, see Appendix \ref{meth:ARI}.

\medskip

\noindent{\it FIR and ARX.}
FIR  \cite{Simpson2001} and ARX models have been applied to CA by several authors \cite{Liu02, Liu03, Panerai03, Nikolic2016}.  First, the raw signals are normalized and detrended. Then, they are fitted using either FIR or  ARX (FIR being a special case of ARX) models of different orders and cerebral autoregulation is estimated by analysing  the phase shift and gain of the corresponding transfer function calculated as the average of the values taken around $[0.07-0.20]$\,Hz. For more details on the computational aspects of FIR and ARX, see Appendix \ref{meth:ARX}.

\medskip

\noindent {\it Mx, Sx and Dx.}
The Mx (Sx and Dx) autoregulatory index is a correlation coefficient between the mean (systolic and diastolic) ABP  and mean CBFV over a certain interval. If a moving window is applied the values are averaged for each investigation and each patient. Several authors \cite{Piechnik1999, Lang2002} have suggested that the values less than 0.3 indicate an intact CA, while  values above 0.3 indicate failure of autoregulation. For more details on the computational aspects of Mx, Sx and Dx, see Appendix \ref{meth:MxSxDx}.

\subsection{Statistical analysis}

\medskip

\noindent {\it Intra-class correlation.} 
An intra-class correlation coefficient (ICC) is used to assess the reproducibility of CA measures, that is their temporal stability during repeated measurements on the same subjects under the same conditions. The ICC is computed as
\begin{equation}\label{eq:ICC}
{\rm ICC} = \frac{\MS_{B}-\MS_{W}}{\MS_{B}+\MS_{W}},
\end{equation}
where $\MS_{B}$, $\MS_{W}$ are between- and within-subjects mean squares \cite{McGraw1996}. Formula \eqref{eq:ICC} quantifies the degree of absolute agreement among measurements; and the values $\MS_{B}$, $\MS_{W}$  can be obtained from the summary tables of one-way ANOVA using each subject as a single factor. Values of ICC close to $1$ indicate high reliability and values close to $0$ no reliability.  We used Fisher's Z-test to assess the significance of the difference between two ICCs 
(here corresponding to the autoregulation indices estimated from a repeated measurements during the sitting and standing protocol).

\medskip

\noindent {\it Coefficient of variation.} 
The variability of the CA indices between subjects is assessed using the coefficient of variation (CoV). It is a relative measure of dispersion computed as the ratio between standard deviation (SD) and the mean value of the autoregulation estimates and expressed as a percentage. Here the 36 measurements taken from the 18 subjects (including the repeated measurements) were concatenated to provide a single value of CoV for both sitting and standing.

\section{Results}

\noindent {\it Data characteristics.} 
Table~\ref{Table:mCV} gives the mean and standard deviation of ABP and CBFV during sitting and standing.   The mean ABP and mean CBFV are lower during the standing protocol, which is associated with the characteristic drop of blood pressure due to orthostatic stress. The SD (understood as the mean of the standard deviations within each subject along time) for ABP is nearly three times greater during standing protocol and nearly twice as large for CBFV. A large increase is not unexpected, given the transient effect seen in Fig. 1. 

\medskip

\noindent {\it CA indices.} 
The mean and standard deviation (calculated across the 36 measurements in each protocol) of the autoregulation indices are shown in Table~\ref{Table:mCA}. Only the difference for the mean values of Mx, ARX(2,2)-gain and ARX(1,5)-phase between the sitting and standing protocol was found to be significant ($p<0.05$, Wilcoxon); they are  marked in Table~\ref{Table:mCA} with an asterisk ($^*$).  To apply the Wilcoxon test we used all the 36 indices  (repeated measurements from 18 subjects) calculated separately for the sitting and standing protocol. 

\smallskip
\noindent {\it Reproducibility.} 
Fig.\,~\ref{figICC} shows ICC for different autoregulation indices and protocols (sitting and standing). For a given autoregulation index and protocol, each ICC was computed using all 36 values (repeated measurements from 18 subjects). For all indices the ICC is greater during the standing protocol, but it is significantly greater ($p<0.05$, Fisher Z-test) only for the correlation measures:  Mx ($p < 0.03$), Sx $(p<0.003)$ and Dx ($p<0.03$), which are marked in Fig.~\ref{figICC} with a square. The values of ICC for the gain and phase of FIR and ARX of all orders  (considered here)  remain relatively constant, especially during sitting.

\smallskip

\noindent {\it Variability.} Fig.\,\ref{figCoV} shows the CoV for different autoregulation indices and protocols.   As in the case of ICC,  each CoV was computed using all 36 values (repeated measurements from 18 subjects). The CoV is greater during standing for the correlation indices, i.e. Mx, Sx, Dx, and the phase of FIR and ARX of all orders; but it is smaller for the ARI and the gain of FIR and ARX. The values of CoV for the gain and phase of FIR and ARX of all orders (considered here) remain relatively constant during the sitting protocol. However, the phase increases with the filter order.

\section{Discussion}

\noindent {\it Main findings.} 
We have performed exploratory work aimed at assessing the reproducibility  and variability of six types of black-box approaches to autoregulation (ARI, Mx, Sx, Dx, FIR and ARX)  during sitting and standing protocols. The results suggest that the data collected upon standing consistently improve the reproducibility of autoregulation parameters. The difference is especially pronounced for Mx, Sx and Dx coefficients, where in this relatively small sample it reaches statistical significance. This is coupled here with an increase in the dispersion of most of the autoregulation indices (Mx, Dx, Sx, and  the phase of FIR and ARX models).

\medskip

\noindent{\it The choice of indices.}
The list of linear, black-box methods to CA studied here is not exhaustive; it does not include, for example,  TFA.  According to the recent guidelines \cite{Claassen2016} the recommended length of ABP-CBFV for TFA should be at least 5\,min long, assuming stationary physiological conditions, which is related with a use of window segments with a minimum length of 100 s to allow sufficient frequency resolution. 

The ARI method followed the original paper by Tiecks et al. \cite{Tiecks95}, however, alternative implementations  have been proposed. In particular Panerai et al. \cite{Panerai03} showed that  a combination of time series modelling (ARMA) and the best fit to one of the ARI step-responses is less susceptible to physiological sources of variability. 

The ARX orders were selected based on the previous studies. Liu and Allen \cite{Liu02} and Liu et al. \cite{Liu03} used ARX(1,5) and Panerai et al. \cite{Panerai03} ARX(2,2). An additional motivation for using ARX(2,2) is that it is a generalization of the difference equations \eqref{mod:ARI_dyn}, which are the basis for the ARI (see Appendix \ref{meth:ARI}).

\medskip

\noindent{\it The choice of data collection protocols.} 
Regardless of the experimental design for data collection It is reasonable to expect that a stronger excitation will lead to a better SNR,  more accurate system identification and better reproducibility of autoregulation measures \cite{Katsogridakis2013, Simpson2004,Liu2005}. The results shown in Fig.~\ref{figICC} support that observation. 
It seems also that not only the strength but also the type of excitation may affect  CA estimates in different ways. Sorond et al. \cite{Sorond2009} showed that both sit-to-stand and thigh-cuff methods lead to a large between-subject variability of ARI, if compared to the baseline data.  Although larger dispersion was associated with the sit-to-stand manoeuvre, it led to a lower intra-subject variability. 

Van Beek et al. \cite{Beek2010} studied the reproducibility of repeated sit-to-stand manoeuvres in 27 healthy elderly subjects using the gain and phase of the TFA. The sit-to-stand procedures led to an increased TFA coherence (compared to spontaneous blood pressure fluctuations), however, it was not coupled with an improved reproducibility (assessed by ICC). There are obvious similarities between the data collection protocol employed by van Beek et al. \cite{Beek2010} and that of Lipsitz et al. \cite{Lip2000} used in the analysis of this paper. In contrast to the conclusions of van Beek et al., our results show that the sit-to-stand manoeuvre improves the reproducibility across autoregulation coefficients. However, in the case of van Beek et al. the interval between the test and retest was significantly longer (3 months). Another difference is that the authors used longer time series data, with repeated sit-to-stand manoeuvres from elderly subjects; so a direct comparison cannot be made.

\medskip

\noindent{\it Sources of variability.} 
A number of factors can affect the variability for CA estimates including the arterial tension of ${\rm PaCO}_2$, autonomic nervous system activity, body temperature, intracranial pressure and intrathoracic pressure \cite{Panerai2014}.  ${\rm PaCO}_2$ is the one of the strongest determinants of CA performance \cite{Panerai1999}. Aaslid et al. \cite{Aaslid1989} showed that hypercapnia leads to a slower response of CBFV to changes in ABP, whilst hypocapnia lead to a more immediate CBFV response.
In \cite{Sorond2009} the authors showed that the sit-to-stand protocol produces a small decline in ${\rm EtCO}_2$, which can potentially alter autoregulation estimates. 

Investigators have used many different time intervals between the repeated measurements, ranging from minutes to days or even months \cite{Brodie2009, Gommer2010, Reinhard2002}. As pointed out in \cite{Panerai2014}, in the context of spontaneous fluctuations of ABP the expectation that reproducibility of CA estimates from repeated measures a long time apart would be poorer is not fully supported by the available studies.

\medskip

\noindent{\it Sympathetic activation.} 
The effect of sympathetic tone on cerebral autoregulation is not well understood and is still debated \cite{Zhang2002, Seifert2011}. Several authors suggested that under normal conditions, the sympathetic activity is minimally involved in autoregulation but when ABP is acutely elevated it might play a protective role by shifting the static autoregulatory curve to the right \cite{Millar1969, Panerai2001, Paulson1989}.

Sorond et al. \cite{Sorond2009} found no evidence for effects of various intensities of sympathetic tone due to thigh-cuffs and sit-to-stand methods on the CBFV response. Panerai et al. \cite{Panerai2001} previously considered eight common manoeuvres, assumed to exhibit different levels of sympathetic drive, but have not been able to detect any significant differences in the CBFV response.

\medskip

\noindent{\it Study limitations.} 
The approximation of cerebral blood flow by the CBFV measured in the MCA is only valid if the diameter of the MCA is constant. 
A more likely source of uncertainty is related with the use of the Finapress monitor to provide continuous estimates of ABP. This is partially alleviated by using the beat-to-beat average instead of raw ABP-CBFV signals. It has also been previously reported \cite{Panerai08} that ARI based on the time series model were comparable with those derived from direct catheter-tip BP measurements in the ascending aorta.

Although there is no consensus regarding the meaningful length of ABP-CBFV time series to calculate the correlation coefficients, including Mx, Dx and Sx (and variety of different intervals have been used, it is believed that only the data collected over a long period of time is clinically significant \cite{Piechnik1999, Reinhard2003}).  

A limitation is related to the relative nature of ICC as a measure of reliability, which stems from equation \eqref{eq:ICC}. If the differences between subjects are small, the ICC values will also be small despite relatively good reproducibility. Similarly, if the differences between subjects are large, so will be the ICC despite relatively poor reproducibility. Thus, whenever ICC is used, it is important to keep in mind that its meaning is restricted to specific populations \cite{Streiner2008, Weir2005}.

We did not present any results related to the goodness of fit of the discussed models as reported by previous studies \cite{ Mitsis2004, Panerai03}. Although this is generally a desirable intermediate feature, it is not in itself the most important performance measure. The main objective is to construct a physiologically relevant measure of CA functionality, in particular capable of distinguishing between the healthy and impaired state \cite{Angarita-Jaimes2014}.  The current work has contributed to this by pointing out possible limitations of the resting protocol in terms of repeatability, and the benefits of greater excitation of blood pressure variations. 

\section*{Acknowledgement}
The authors acknowledge the support of the EPSRC project EP/K036157/1. The data comes Dr. L. Lipsitz lab and has been previously published \cite{Lip2000}. 

{\small
\setstretch{0.8}

}

\appendix
\section{Computational details}

\subsection{ARI}\label{meth:ARI}
The method follows Tiecks et al. \cite{Tiecks95}. The beat-to-beat average of ABP and CBFV have been obtained as described in Section~\ref{data}, and denoted by $P[k]$ and $V[k]$, respectively. The signals are indexed by $k$ and are of length $N$. Let $\bar P$ and $\bar V$ denote the mean values of $P[k]$ and $V[k]$ for the entire interval of interest. Initially, the time-varying ABP signal $P[k]$ is normalized as follows
\[
\dP[k]=\frac{P[k]-\bar P}{\bar P-P_{cr}},
\]
where $P_{cr}=12\,\mathrm{mmHg}$ is the critical closing pressure \cite{Tiecks95}. The following system of difference equations is used to compute the intermediate quantities $x_1[k]$ and $x_2[k]$ as
\begin{equation}\label{mod:ARI_dyn}
\begin{aligned}
&x_1[k]=x_1[k-1]+\frac{\dP[k]-x_2[k-1]}{fT}\\
&x_2[k]=x_2[k-1]+\frac{x_1[k-1]-2Dx_2[k-1]}{fT},
\end{aligned}
\end{equation}
where $f$, $D$ and $T$ are the sampling frequency, damping factor and time constant parameters, respectively. The modelled beat-to-beat average of CBFV, denoted by $\hat V[t]$, is computed as
\begin{equation}\label{mod:ARI_V}
\hat V[k]=\bar V(1+\dP[k]-Kx_2[k]),
\end{equation}
where $K$ is a parameter reflecting autoregulatory gain. In order to start the process at the dynamics close to the baseline, the initial conditions were selected as
\begin{equation}\label{ARI:ic}
x_1(0) = 2D\,\dP[0],\qquad x_2(0) = \dP[0].
\end{equation}

\begin{figure}[h!]
\begin{center}
\includegraphics[width=0.85\textwidth]{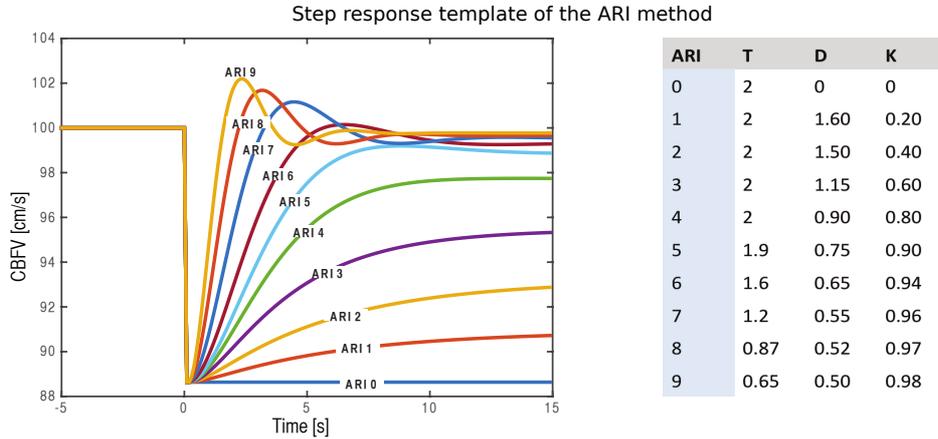}
\end{center}
\caption{The ten levels of ARI based on the ABP-CBFV response of the Aaslid-Tiecks model with the corresponding values of parameters 
(T,D,K) provided in the table. }\label{figARI}
\end{figure}

\begin{figure}[h!]
\begin{center}
\includegraphics[width=1\textwidth]{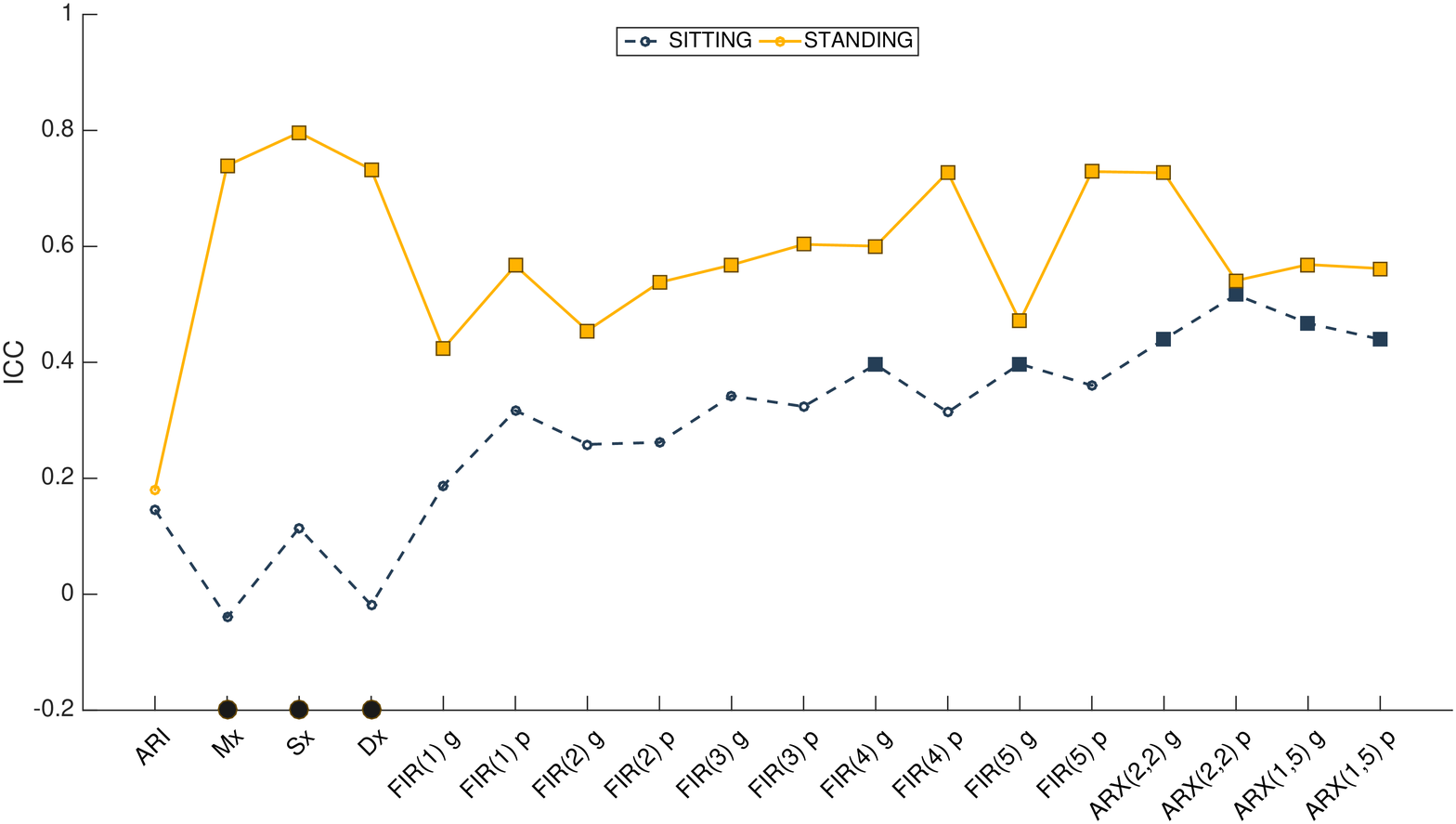}
\end{center}
\caption{The ICC values for different CA indices during sitting (dashed blue) and standing (solid yellow) protocol. The indices with the significant ($p<0.05$) population value of ICC different than zero are marked with squares; and indices with a significant difference between sitting and standing are denoted by black circles.  FIR(n) g and FIR(n) p denotes the gain and phase, respectively, of the FIR model of the order n. Similar notation is used for the indices derived from the ARX model.}\label{figICC}
\end{figure}

\begin{figure}[h!]
\begin{center}
\includegraphics[width=1\textwidth]{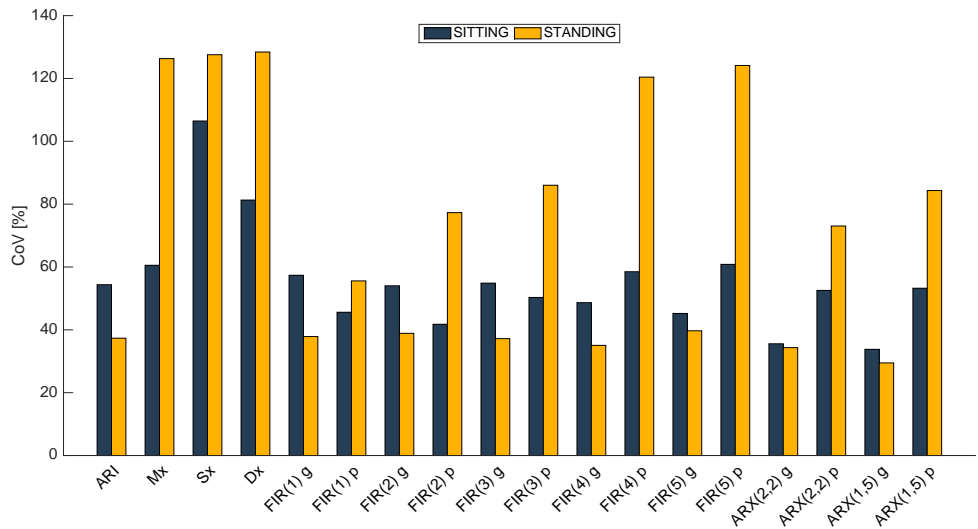}
\end{center}
\caption{Within-group variability (CoV) of CA measures during sitting (blue) and standing (yellow) protocols. FIR(n) g and FIR(n) p denotes the gain and phase, respectively, of the FIR model of the order n. Similar notation is used for the indices derived from the ARX model.}\label{figCoV}
\end{figure}

In \cite[Table\,3]{Tiecks95} combinations of ten different values of $(T,D,K)$ were used to generate ten models corresponding to various grades of autoregulation ranging from  0 (absence of autoregulation) to 9 (strongest autoregulation). Let $\hat V_j[k]$ denote the response \eqref{mod:ARI_V} of the model for the $j$th ($j=0,\ldots,9$) combination of the parameters $(T,D,K)$ given in Fig.~\ref{figARI}. The difference between the predicted and measured CBFV is computed as  $d_j =  \|(\hat V_j[k] - V[k])/\bar V\|$, where $\|\cdot \|$ is the $\ell^2$-norm.  By $f_{\ari}(s)$ we denote the interpolation by cubic splines of the values $d_j$. Finally, the \ari\,\,is determined by calculating the $s\in[0,9]$ that minimizes $f_{\ari}(s)$, that is
\begin{equation}\label{ARI}
\ari = \argmin_{s\in[0,\ldots 9]} f_{\ari}(s).
\end{equation}

\subsection{FIR and ARX}\label{meth:ARX}
The method  follows \cite{Liu02,Liu03}. The time series $P[k]$ and $V[k]$ are normalized about their mean values
\begin{equation}\label{nd}
\nP[k] = \frac{P[k] - \bar P}{\bar P},\qquad \nV[k] = \frac{V[k]-\bar V}{\bar V}, \end{equation}
and then linearly detrended. The method relates the input and output by the linear difference model
\begin{equation}\label{ARX:mod}
\nV[k]  = - \sum_{j=1}^{n_a} a_j \nV[k-j] +  \sum_{j=0}^{n_b} b_j \nP[k-j]+e[k],
\end{equation}
where $a_j$ and $b_j$ are the parameters;  $n_a$, $n_b$ are the model orders; $e[k]$ is the model error. Note that model \eqref{ARX:mod} has an easy interpretation: the output $\nV[k]$ is determined as a linear combination of its $n_a$ previous values and $n_b$ consecutive values of the input $\nP[k]$. 
The resulting model will be denoted by ${\rm ARX}(n_a,n_b)$. Note that by definition the ${\rm FIR}(m)$ model will be simply ${\rm ARX}(m,0)$.

\smallskip

The estimated model parameters $a_j$, $b_j$ define the transfer function
\begin{equation}\label{tf}
{\rm H}(z) = \frac{b_0+b_1 z^{-1}\ldots b_{n_b} z^{-n_b}}{a_0+a_1 z^{-1}\ldots a_{n_a} z^{-n_a}}
\end{equation}
associated with the discrete model \eqref{ARX:mod}, where due to normalization $a_0=1$. The average phase and gain   of  ARX  is typically reported, that is
\[
{\rm ARX\,\, gain} = \frac{1}{M} \sum_{k=1}^M |H(e^{j\omega_kT_s})|\quad\qquad {\rm ARX\,\, phase} = \frac{180}{M \pi} \sum_{k=1}^{M} \angle[H(e^{j\omega_kT_s})],
\]
where $T_s$ is the sampling frequency; $\omega_k$ are the $M$ frequencies at which the response is computed; $|z|$ and $\angle(z)$ denotes the absolute value and the angle of the complex number $z$, respectively.  
Since the power spectra of ABP and CBFV both show a peak in the low frequency band at approximately $0.1\, {\rm Hz}$ (see \cite{Lip2000}), the average phase and gain are estimated in the frequency range $[0.07-0.20]\,{\rm Hz}$ \cite{Zhang98, Panerai03}.

\subsection{Mx, Sx and Dx}\label{meth:MxSxDx}
As before, let $P[k]$ and $V[k]$ denote the beat-to-beat average of ABP and CBFV, respectively, of length $N\in\N$, as described in Section~\ref{data}.  The Mx index is computed as Pearson's correlation coefficient
\begin{equation}\label{Mx}
Mx = \frac{\sum_{k=1}^{N}(P[k]-\bar P)(V[k]-\bar V)}{\sqrt{\sum_{k=1}^{N}(P[k]-\bar P)}\sqrt{\sum_{k=1}^{N}(V[k]-\bar V)}},
\end{equation}
where $\bar P$ and $\bar V$ are the mean of the time series $P[k]$ and $V[k]$, respectively taken over the interval under consideration. Similarly, the Sx and Dx indices are computed using \eqref{Mx} with the beat-to-beat average $P[k]$ and mean $\bar P$  of ABP replaced with beat-to-beat average of systolic and diastolic pressure.

\clearpage
\section{Tables and Graphs}

\begin{table}[h!]
\centering
\begin{tabular}{cccccc}
\hline
			&          		& \multicolumn{2}{c}{Sitting} & \multicolumn{2}{c}{ Standing}\\
{ Quantity} 	& Units 		& Mean         	& SD 	      	& Mean 		 	& SD \\
\hline
ABP 			& mmHg 	    	& 88.43        	& 2.875 	      	& 85.44        		& 8.374 \\
CBFV		& cm/s 	    	& 42.74 	   	& 2.268 	      	& 40.35		 	& 4.129 \\
\hline
\end{tabular}
\caption{The main characteristics of the beat-to-beat average ABP and CBFV. The `Mean' is the calculated as the mean across all subjects for sitting (baseline) and standing (transient) data segments; and `SD' is the mean of the standard deviations within each subject along time.}\label{Table:mCV}
\end{table}

\medskip

\begin{table}
\centering
\begin{tabular}{lcccc}
\hline
			    & \multicolumn{2}{c}{Sitting}			& \multicolumn{2}{c}{Standing} \\
{Method} 		& Mean 		    					& SD 		    	& Mean 		    					& SD \\
\hline
ARI	 			& 4.97	  	    				& 2.71 		    	& 5.68 		    					& 2.12\\
Mx  				& \hspace{1ex}0.44$^{*}$ 			& 0.27		    	& \hspace{1ex}0.30$^*$   				& 0.38\\
Sx  				& 0.28		    				& 0.30		    	& 0.30		    					& 0.38\\
Dx  				& 0.38		    				& 0.31		    	& 0.31		    					& 0.39\\
{\rm FIR(1)}		& [1.56, \hspace{1ex}47.3]		& [0.89, 21.6]		& [1.46, \hspace{1ex}46.1]			& [0.55, 25.6]\\
{\rm FIR}(2) 		& [1.82, \hspace{1ex}41.3]		& [0.98, 17.3]		& [1.60, \hspace{1ex}33.8]			& [0.62, 26.1]\\
{\rm FIR}(3) 		& [1.86, \hspace{1ex}38.3]		& [1.02, 19.3]		& [1.58, \hspace{1ex}31.9]			& [0.59, 27.5]\\
{\rm FIR}(4) 		& [1.81, \hspace{1ex}37.7]		& [0.88, 22.1]		& [1.59, \hspace{1ex}26.8]			& [0.56, 32.3]\\
{\rm FIR}(5) 		& [1.77, \hspace{1ex}38.4]		& [0.80, 23.4]		& [1.60, \hspace{1ex}25.9]			& [0.63, 32.1]\\
{\rm ARX}(2,2) 		&[1.74$^*$, 37.6]				& [0.62, 19.7]		& [1.49$^*$, 30.0]					& [0.51, 21.9]\\
{\rm ARX}(1,5) 		&[1.60, 40.7$^*$]				& [0.54, 21.7]		& [1.59, 30.3$^*$]					& [0.47, 25.5]\\
\hline	
\end{tabular}
\caption{CA measures. Group mean and SD calculated from 18 normotensive subjects for the sitting and standing protocol. The values given in brackets for ${\rm FIR}(n_b)$ and ${\rm ARX}(n_a,n_b)$ correspond to the mean and SD of the gain [$\%/\%$]  and phase [$^\circ$]  estimated in the low frequency range.) The significant difference ($p<0.05$, Wilcoxon) of the autoregulation indices between sitting and standing is denoted with an asterisk ($^*$).}
\label{Table:mCA}
\end{table}


\end{document}